\documentclass[10pt]{IEEEtran}

\usepackage[cmex10]{amsmath}
\usepackage{amsfonts}
\usepackage[font=footnotesize]{caption}
\usepackage{cite}
\usepackage{amsthm}
\usepackage{amssymb}
\usepackage{circuitikz}
\usepackage{tikz}
\usepackage{subfig}    
\usepackage{float}
\usetikzlibrary{hobby}
\usepackage[normalem]{ulem}
\usepackage{pgfplots}
\usepackage{pifont}
\usepackage[colorinlistoftodos,prependcaption,textsize=small]{todonotes}
\usepackage{regexpatch}
\makeatletter
\xpatchcmd{\@todo}{\setkeys{todonotes}{#1}}{\setkeys{todonotes}{inline,#1}}{}{}
\makeatother

\begin{document}
\title{Synthesis of General Decoupling Networks Using Transmission Lines}
\author{Binbin Yang
	\thanks{ 1 Binbin Yang is with the Department of Electrical and Computer Engineering, North Carolina Agricultural and Technical State University, Greensboro, NC, USA (e-mail: byang1@ncat.edu).   }
}


\maketitle

\begin{abstract}
In this paper, we introduce a synthesis technique for transmission line based decoupling networks, which find application in coupled systems such as multiple-antenna systems and compact antenna arrays. Employing the generalized $\pi$-network and the transmission line analysis technique, we reduce the decoupling network design into simple matrix calculations. The synthesized decoupling network is essentially a generalized $\pi$-network with transmission lines at all branches. A standard electrical length of $3\lambda/8$ and $5\lambda/8$ are chosen to simplify the physical implementation, leaving the characteristic impedances of the transmission line branches the main design parameters. The advantage of this proposed decoupling network is that it can be implemented using transmission lines, ensuring better control on loss, performance consistency and higher power handling capability when compared with lumped components, and can be easily scaled for operation at different frequencies. A two-port microstrip antenna system at 1.2 GHz and a three-port monopole antenna system at 1 GHz are investigated respectively to demonstrate the validity of the proposed synthesis method, and perfect decoupling ($S_{21}<-50$dB) are achieved at both design frequencies. 

\end{abstract}


\section{Introduction}
\IEEEPARstart{A}{ntenna} arrays and multiple-antenna systems typically suffer from mutual coupling effect, which causes lower system efficiency, impedance mis-matching \cite{balanis2016antenna} and correlation-induced information capacity reduction \cite{wallace2004mutual}. Therefore, a decoupling network (DN) that combats the mutual coupling effect is of great interest to such coupled systems. The reported works in literature on decoupling techniques typically deal with coupled networks with special characteristics, such as circulant symmetry \cite{coetzee2009design}, and systems with limited number of ports \cite{lau2012simple,wu2015very,xia2014efficient,li2019novel,li2020novel}. For example, the work in \cite{li2020novel,xia2014efficient,Efficient2019zou} consider only two-port systems and introduce special transmission line structures for the DN design \cite{xia2014efficient,Efficient2019zou}. The work in \cite{lau2012simple} introduces a parasitic antenna element as extra degrees of freedom for decoupling, but only applies to two-port systems as well. The networks in \cite{coetzee2009design} though tackle multi-port systems, only apply to networks with circular symmetry. 

The first few general decoupling networks, to the best of the author's knowledge, are reported in \cite{nie2014systematic,geren1996practical,volmer2008eigen,yang2020decoupling}. In \cite{nie2014systematic}, a decoupling network using lumped elements is proposed based on the analysis of generalized LC $\pi$-network. However, its implementation using lumped elements suffers from parasitic effects and losses in lumped elements. Moreover, in real application, extra physical layout traces are needed and will add extra transmission line effects to the DN's response, potentially deviating the DN's performance from the synthesized optimal case, especially at high frequencies. In \cite{geren1996practical}, multi-port decoupling networks are synthesized as cascaded directional couplers, which are then implemented using lumped components, and suffer from similar shortcomings at high frequencies as in \cite{nie2014systematic}. The work in \cite{volmer2008eigen} also proposes a scheme for decoupling multi-port radiators with an implementation using 180$^\circ$ directional couplers. In \cite{yang2020decoupling}, a decoupling network using transformer banks is proposed based on characteristic port modes, which uses the minimum number of components for general decoupling and matching, but its implementation requires efficient transformers at radio and microwave frequency ranges. There is clearly a lack of synthesis technique for general decoupling networks using purely transmission lines, which exhibit higher energy efficiency, easier control on network performance, and simple scalability at various frequencies.

\begin{figure}[t]
		\centering
        \includegraphics[width=0.99\linewidth]{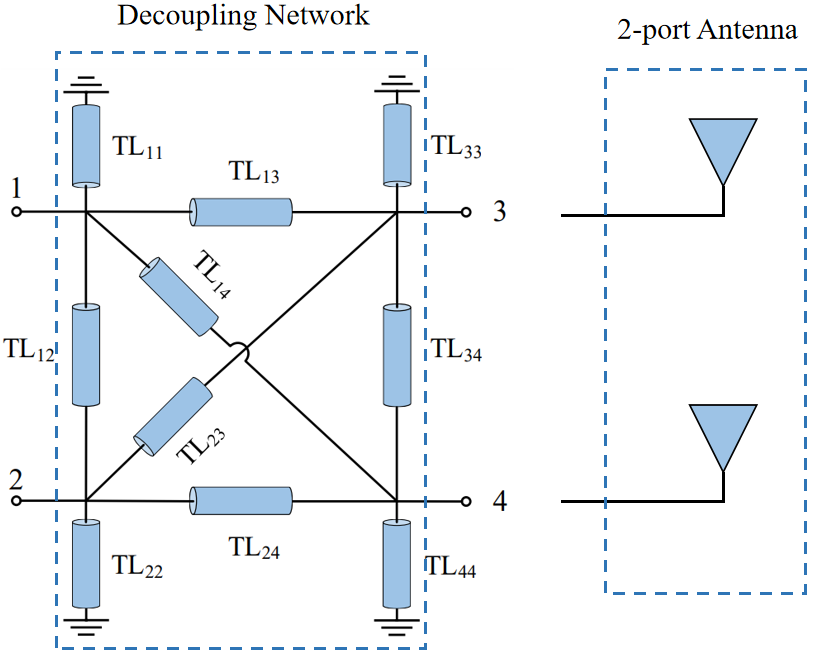}%
        \caption{The topology of the generalized TL $\pi$ network (two-port case). Ports 3 and 4 can be connected to a coupled two-port system, and ports 1 and 2 provide access to a decoupled two-port response.}\label{fig:TL_pi}
\end{figure}

In \cite{sinha2016synthesis}, a network synthesis technique using transmission lines is proposed. Though generic, the paper \cite{sinha2016synthesis} fails to propose a network topology that could synthesize arbitrary networks, and only conventional ring hybrid and directional couplers are considered in the given examples. Moreover, its analysis technique relies mainly on ABCD parameters, making it quite cumbersome for different systems with varying number of ports, as new equations has to be derived once the number of ports are changed. 

In this paper, we will demonstrate that a transmission line (TL) based generalized $\pi$-network can be used to synthesize an arbitrary passive, reciprocal and lossless decoupling network. The analysis of the proposed network is based on Y parameters, which significantly simplifies the DN synthesis process for the proposed topology. The same set of equations can be scaled to analyze networks with arbitrary number of ports, as demonstrated in later sections through a two-port and a three-port examples.

\section{Theory of the TL-based Decoupling Network}
\label{section:Methodology}
\label{section:math_TL_DN}
The proposed DN is based on a generalized $\pi$-network of transmission lines as shown in Fig. \ref{fig:TL_pi} for a two-port antenna system. Each branch of the network represents a transmission line. In this case, two of the ports (e.g. ports 3 and 4) can be connected to a coupled two-port system, and the other two ports (e.g. ports 1 and 2) provide access to a decoupled two-port response (assuming the TL branches are designed accordingly). For an N-port coupled system, the DN will be a 2N-port network, with similar generalized $\pi$-network configuration as in Fig. \ref{fig:TL_pi}. Namely, between each two ports, there will be a TL branch connecting them, and at each port, one TL branch is shorted to ground, constituting a total of $2N^2+N$ branches. Half of the ports will be connected to an N-port antenna system, with the other N ports providing decoupled responses. 

Following is the detailed theory for general multi-port decoupling network synthesis using transmission lines. First, consider a uniform transmission line with the characteristic impedance $Z_0$ and the electrical length $\theta$. Its ABCD parameters can be represented as  
\begin{equation}
\label{eq:ABCD_TL}
A_{TL}=\begin{bmatrix}
\cos\theta & jZ_0\sin\theta\\ 
jY_0 \sin\theta & \cos\theta
\end{bmatrix}=\begin{bmatrix}
a & jb\\ 
jc & a
\end{bmatrix},
\end{equation}
where $a=\cos\theta$ and $b=Z_0\sin\theta$. It can be checked that $a^2+bc=1$ for reciprocal networks.

The admittance Y parameter of one TL branch can be obtained through conversion from the ABCD parameters in \eqref{eq:ABCD_TL} as \cite{frickey1994conversions}
\begin{equation}
\label{eq:Y_2port}
y_{TL}=\begin{bmatrix}
\frac{a}{jb} & \frac{-1}{jb}\\ 
\frac{-1}{jb} & \frac{a}{jb}
\end{bmatrix}
\end{equation}

Now consider a 2N port generalized $\pi$ network. For each two ports (e.g. ports $n$ and $k$), the corresponding Y matrix for the transmission line between ports $n$ and $k$ can be represented as 
\begin{equation}
\label{eq:Ynk_2port}
y_{n,k}=\begin{bmatrix}
y_{n,k}^{11} & y_{n,k}^{12}\\ 
y_{n,k}^{21} & y_{n,k}^{22}
\end{bmatrix}
=\begin{bmatrix}
\frac{a_{n,k}}{jb_{n,k}} & \frac{-1}{jb_{n,k}}\\ 
\frac{-1}{jb_{n,k}} & \frac{a_{n,k}}{jb_{n,k}}
\end{bmatrix}
\end{equation}
and at each port (e.g. port $n$), there is a TL branch shorted to ground with its Y parameter represented as
\begin{equation}
\label{eq:Ynn_2port}
y_{n,n}=y_{n,n}^{11}=\frac{a_{n,n}}{jb_{n,n}}
\end{equation}

Following the analysis in \cite{nie2014systematic}, the Y parameter of a 2N-port generalized $\pi$-network can be assembled from the Y parameters of each branch as 
\begin{equation}
Y_{\Pi}=
\begin{bmatrix}
    \sum_{k=1}^{2N}y_{1,k}^{11}     & y_{1,2}^{12} & y_{1,3}^{12} & \dots  & y_{1,2N}^{12} \\
    y_{2,1}^{12}    & \sum_{k=1}^{2N}y_{2,k}^{11} & y_{2,3}^{12}& \dots  & y_{2,2N}^{12}\\
    y_{3,1}^{12}    & y_{3,2}^{12}& \sum_{k=1}^{2N}y_{3,k}^{11} &  \dots  & y_{3,2N}^{12}\\
    \vdots & \vdots & \vdots & \ddots & \vdots \\
    y_{2N,1}^{12} & y_{2N,2}^{12} &y_{2N,3}^{12}& \dots  & \sum_{k=1}^{2N}y_{2N,k}^{11}
\end{bmatrix}
\label{eq:Y_pi}
\end{equation}
or more explicitly
\begin{equation}
Y_{\Pi}=
\begin{bmatrix}
    \sum_{k=1}^{2N}\frac{a_{1,k}}{jb_{1,k}}     & \frac{-1}{jb_{1,2}} & \frac{-1}{jb_{1,3}} & \dots  & \frac{-1}{jb_{1,2N}} \\
    \frac{-1}{jb_{2,1}}    & \sum_{k=1}^{2N}\frac{a_{2,k}}{jb_{2,k}} & \frac{-1}{jb_{2,3}}& \dots  & \frac{-1}{jb_{2,2N}}\\
    \frac{-1}{jb_{3,1}}    & \frac{-1}{jb_{3,2}}& \sum_{k=1}^{2N}\frac{a_{3,k}}{jb_{3,k}} &  \dots  & \frac{-1}{jb_{3,2N}}\\
    \vdots & \vdots & \vdots & \ddots & \vdots \\
    \frac{-1}{jb_{2N,1}} & \frac{-1}{jb_{2N,2}} &\frac{-1}{jb_{2N,3}}& \dots  & \sum_{k=1}^{2N}\frac{a_{2N,k}}{jb_{2N,k}}
\end{bmatrix}
\label{eq:Y_pi}
\end{equation}
where $b_{n,k}=b_{k,n}$ and $a_{n,k}=a_{k,n}$ due to the reciprocal property of each transmission line branch. One way to understand the matrix assembly is to consider the definition of Y parameters, which assumes the inactive ports being shorted. For example, $Y^{11}$ of a multiport network is defined when all the other ports are shorted to ground, leaving only port 1 being active. For the generalized $\pi$ network, all the ports being shorted to ground (except for port 1), leaves only the $y_{1,k}^{11}$ components in the response, hence $Y^{11}=\sum_{k=1}^{2N}y_{1,k}^{11}$. The mutual terms following similar analysis turn out to remain the same as $y_{1,k}^{12}$ for ports $1$ and $k$.

In the proposed TL-decoupling network, the number of unknowns to be decided is $(4N^2+2N)$ (two unknowns $a$ and $b$, or $Z_0$ and $\theta$ for each TL branch), which is twice as many as what is needed to synthesize an arbitrary reciprocal and lossless 2N-port network \cite{nie2014systematic}. 
By enforcing the characteristic impedance (or the electrical length) on each branch, the number of unknowns is reduced to $2N^2+N$. Now enforcing a conjugate matching with the $N-$port load, the number of unknowns is reduced to $N^2$. As discussed in \cite{nie2014systematic}, this $N^2$ design freedom implies that there is no unique solution for the multi-port DN. This fact can be leveraged to simplify the design topology or optimize the matching bandwidth of the DN.

\subsection{DN Synthesis Process}
\label{sec:DN_synthesis}
Before synthesizing the TL DN for a specific multi-port load, a mathematical DN that achieves the decoupling for the given load is needed, which can be obtained using different methods \cite{nie2014systematic,domizioli2012front}. We here adopt the decoupling network derived from S parameters given in \cite{nie2014systematic}. For a given N-port load with an S parameter of $S_L$, it's convenient to get its singular value decomposition form as $S_L=U_L^H \Lambda_L V_L$. The corresponding S matrix for a 2N-port decoupling network that decouples the given N-port load can be then represented as \cite{nie2014systematic}:
\begin{equation}
\label{eq:svd_DN}
S_{DN}=\begin{bmatrix} -VV_L^\mathrm{H} U_L^*\Lambda_L V^T & V(I-\Lambda_L^2)^\frac{1}{2} U_L^\mathrm{H} \\ U_L^*(I-\Lambda_L^2)^\frac{1}{2} V^T & V_L\Lambda_L U_L^\mathrm{H} \end{bmatrix}
\end{equation}
As noted in \cite{nie2014systematic}, the unitary matrix $V$ is random, and represents the $N^2$ design freedom. Each unitary matrix $V$ corresponds to an $S_{DN}$. The DN's admittance matrix $Y_{DN}$ can be obtained through direct S to Y conversion as
\begin{equation}
\label{eq:S2Y_DN}
Y_{DN}=Y_0(I-S_{DN})(I+S_{DN})^{-1}
\end{equation}
where $Y_0=1/Z_0$. 

Since $Y_{DN}$ is determined once the load $S_L$ and the unitary matrix $V$ are specified, the synthesis of the DN is now reduced to solving the following equation for all the $a$ and $b$ values in \eqref{eq:Y_pi} from the following mapping relation
\begin{equation}
\label{eq:Y_pi_Y_DN}
Y_\Pi=Y_{DN}
\end{equation}

To ensure a TL implementation, we first specify all the $a$'s in $Y_{DN}$ in \eqref{eq:Y_pi} to be some value in the range of $-1$ and $1$ (note $a=\cos\theta$), and then solve equation \eqref{eq:Y_pi_Y_DN} to get the solution of the remaining unknowns (all the $b$'s).

After finding the values of $a$ and $b$ on each TL branch, we can convert them to the electrical length and characteristic impedance of the TL branch using the following relation

\begin{equation}
\label{eq:theta}
\theta=\cos^{-1}a
\end{equation}
\begin{equation}
\label{eq:Z0}
Z_0=b/\sin\theta
\end{equation}

Though it is always desirable to have a small $\theta$ to make the TL shorter, sometimes the resulted $b$ being negative requires that $\sin\theta$ has to be negative. In this case, the $\theta$ derived from \eqref{eq:theta} has to be modified as $2\pi-\cos^{-1}a$ to meet the requirement. Namely, 
\begin{equation}
\label{eq:theta_2}
\theta=2\pi-\cos^{-1}a \;\; \mathrm{if}\;\; b<0
\end{equation}

\begin{figure}[t]
		\centering
        \subfloat[]{\includegraphics[width=0.65\linewidth]{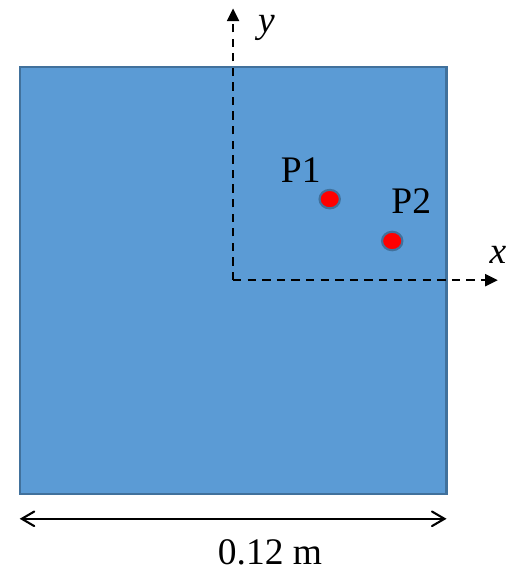}%
        \label{fig_first_case}}
                \hfil
        \subfloat[]{\includegraphics[width=0.8\linewidth]{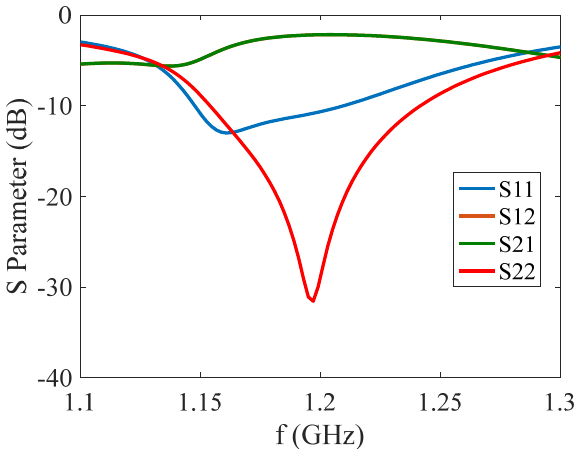}%
        \label{fig_second_case}}
                        \hfil
        \subfloat[]{\includegraphics[width=0.8\linewidth]{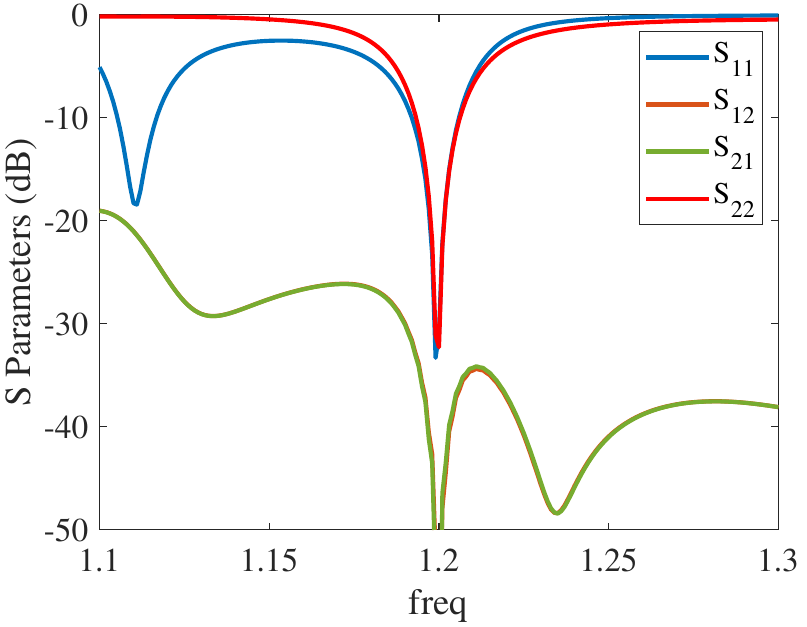}%
        \label{fig_second_case}}
                  
        \caption{(a) the geometry of the two-port square patch antenna, (b) the original S parameter behavior of the two-port patch, (c) the S parameter of the two-port patch with DN synthesized using standard electrical length ($S_{12}$ overlaps with $S_{21}$).}\label{fig:patch_2ports}
\end{figure}

\subsection{DN Synthesis with Standard Electrical Length}
\label{sec:standard_theory}
As stated above, the TL-based DN has double the number of unknowns compared with the LC DN network. To reduce the number of unknowns and standardize the DN implementation, we here fix certain parameters. Specifically, $a=\cos\theta$ is set as $-1/\sqrt{2}$, namely a standard electrical length of $3\lambda/8$ or $5\lambda/8$ long. The parameters to be determined for the DN design are then the characteristic impedance $Z_0$ (or $b$'s) for each branch TL. The resulting ABCD parameter of the TL section is 

\begin{equation}
A_{TL}=\frac{-1}{\sqrt{2}}\begin{bmatrix}
1 & jZ_0\tan\theta\\ 
jY_0 \tan\theta & 1
\end{bmatrix}
\end{equation}
where $a=\frac{-1}{\sqrt{2}}$, and $b=-Z_0\tan\theta/\sqrt{2}$. 

With the $a$'s being specified, the $b$ values can be determined by solving equation \eqref{eq:Y_pi_Y_DN} (see the analysis in Section \ref{sec:DN_synthesis}). Then, the length and the characteristic impedance of each TL branch can be determined from the $b$'s. In this case, the electrical length and the $Z_0$ of each TL branch depends on the $b$ value as 

\begin{equation}
\label{eq:length_standard}
  l =
    \begin{cases}
      3\lambda/8 & \text{if $b>0$}\\
      5\lambda/8 & \text{if $b<0$}
    \end{cases}  
\end{equation}

\begin{equation}
\label{eq:Z0_standard}
  Z_0 = \sqrt{2}|b|
\end{equation}

It's worth noting that the choice of $l$ could also be $\lambda/8$ and $7\lambda/8$ if $\cos\theta=1/\sqrt{2}$. We here select the relation in \eqref{eq:length_standard} in consideration of easier layout with comparable TL lengths.

\section{Design Examples}
Two numerical examples are examined here to validate the proposed DN. To avoid randomness and facilitate crosscheck, we pick the matrix $V$ in \eqref{eq:svd_DN} as an identity matrix $I$ when determining our $S_{DN}$.

\subsection{Two-port Antenna System}
The first example is a two-port patch antenna. As shown in Figure \ref{fig:patch_2ports} (a), the square patch has two ports arbitrarily placed at P1 (0.025, 0.021) and P2 (0.044, 0.009). The patch has a size of $0.12\times0.12$ m$^2$, and is placed 5 mm above the ground plane with an air substrate. The frequency response of the antenna before decoupling is shown in Figure \ref{fig:patch_2ports} (b). Strong mutual coupling is clearly observed between the two ports across the entire observed frequency range.

We now seek to design a DN to decouple the two-port antennas at 1.2 GHz. Following the analysis in Section \ref{section:math_TL_DN}, we first find out the S parameter of the two-port antennas at 1.2 GHz. Then the DN S paramter $S_{DN}$ is arrived at through \eqref{eq:svd_DN} assuming $V=I$. The DN can then be designed by solving \eqref{eq:Y_pi_Y_DN} for $a$'s and $b$'s. The $a$'s and $b$'s are then mapped to electrical length and characteristic impedances using relations \eqref{eq:length_standard} and \eqref{eq:Z0_standard}. Standard electrical length is used here for design simplification. 
Table \ref{tab:standard} shows the corresponding electrical length and characteristic impedance of each TL branch for the DN. Figure \ref{fig:patch_2ports} (c) shows the S parameter of the two-port antenna after inserting the synthesized TL-based DN in front of the coupled two-port antennas. As expected, the two ports are perfectly decoupled at the design frequency 1.2 GHz ($S_{21}<-50$dB). The DN at a different frequency can be designed following similar procedure.

\begin{table}
\renewcommand{\arraystretch}{1.5}
\captionsetup{justification=centering}
\caption{Decoupling Network Parameters for the Two-Port System with Standard Electrical Length} 
\centering 
\begin{tabular}{|c |c | c |} 
\hline
TL Branch & $Z_0$ ($\Omega$) & $\theta$ (degree)\\
\hline 
$TL_{11}$ & 109.71 & 225\\
\hline 
$TL_{12}$ & 15425.54 & 225\\
\hline 
$TL_{13}$ & 196.00 & 135\\
\hline 
$TL_{14}$ & 239.55 & 225\\
\hline 
$TL_{22}$ & 88.71 & 135\\
\hline 
$TL_{23}$ & 127.90 & 135\\
\hline 
$TL_{24}$ & 100.43 & 135\\
\hline 
$TL_{33}$ & 42.35 & 135\\
\hline 
$TL_{34}$ & 69.50 & 225\\
\hline 
$TL_{44}$ & 43.51 & 135\\
\hline 
\end{tabular}
\label{tab:standard} 
\end{table}

\begin{figure}
		\centering
        \subfloat[]{\includegraphics[width=0.8\linewidth]{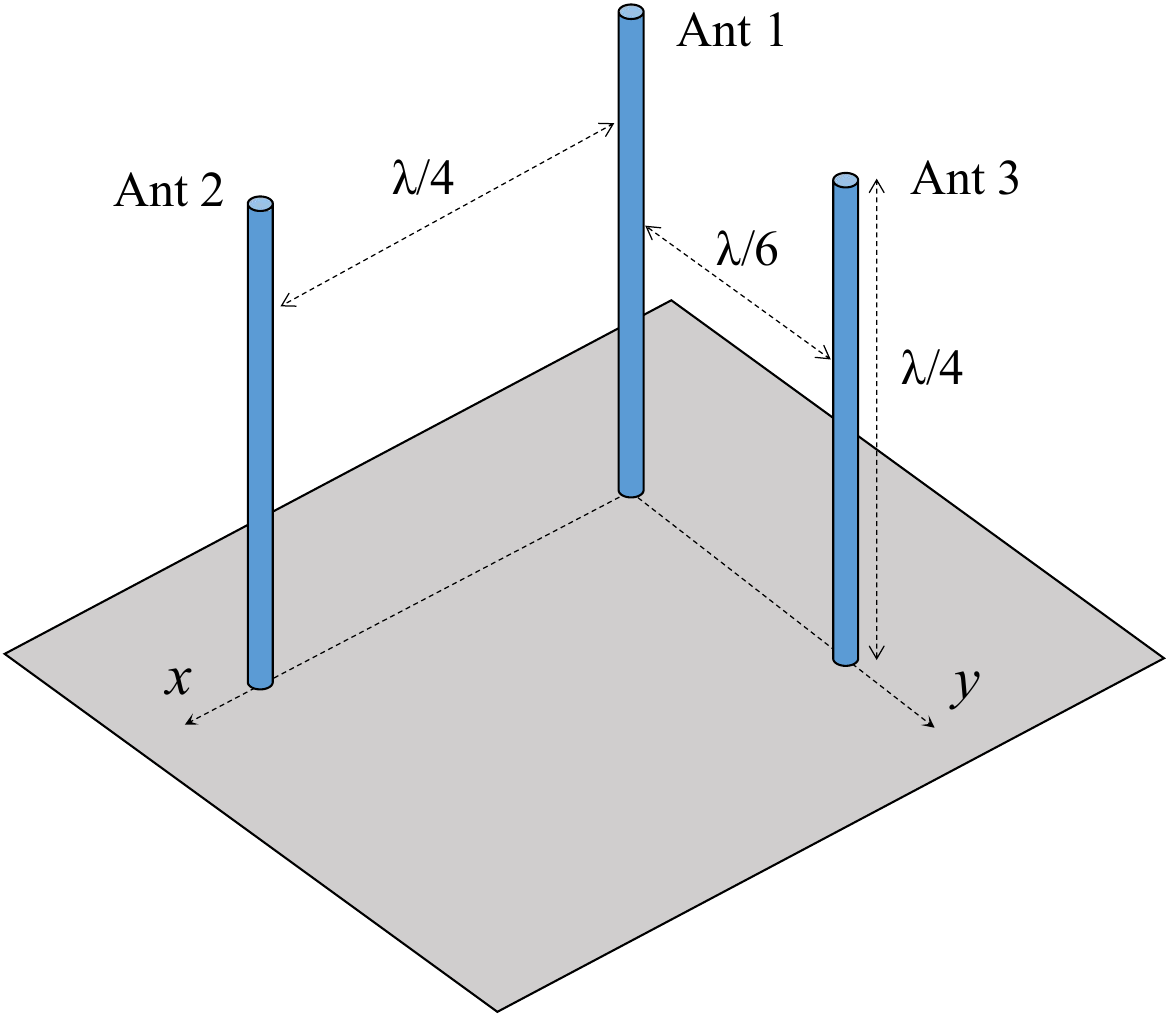}
        \label{fig_first_case}}
                \hfil
        \subfloat[]{\includegraphics[width=0.85\linewidth]{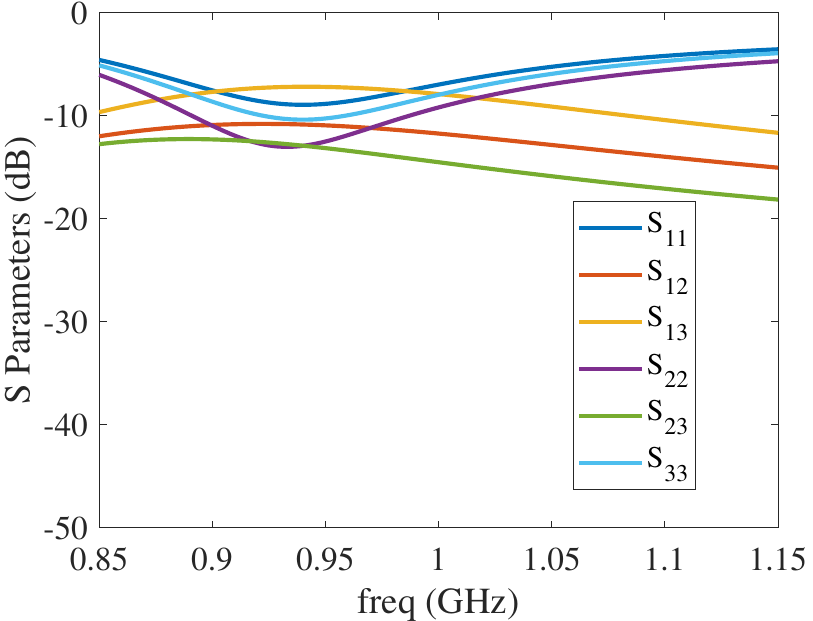}%
        \label{fig_second_case}}
                        \hfil
        \subfloat[]{\includegraphics[width=0.85\linewidth]{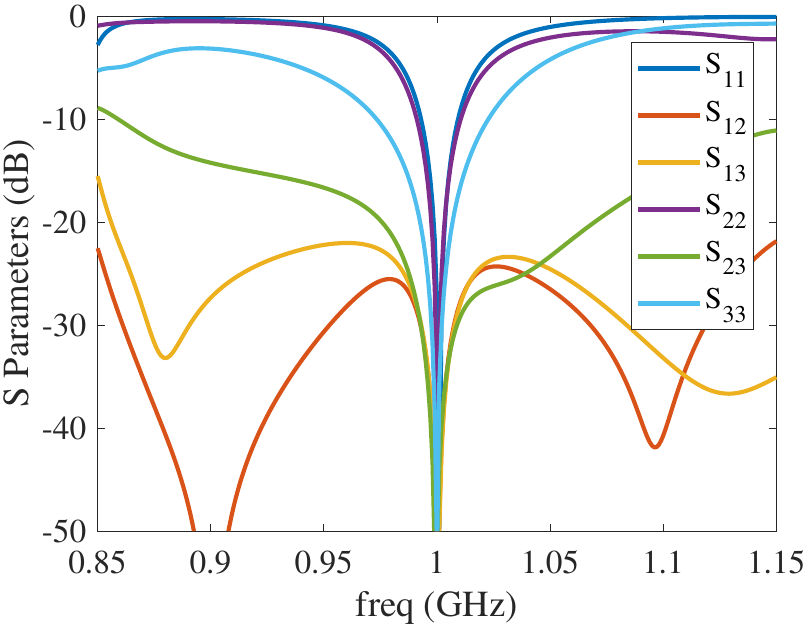}%
        \label{fig_second_case}}
                  
        \caption{(a) the geometry of the three monopoles, (b) the original S parameter behavior of the three monopoles, (c) the S parameter of the three monopoles with DN synthesized using standard electrical length.}\label{fig:monopole_3ports}
\end{figure}

\subsection{Three-port Antenna System}
Here a three-port antenna system based on strongly coupled monopoles operating at 1 GHz is used for the DN test. As shown in Figure \ref{fig:monopole_3ports} (a), the monopoles are quarter wavelength long (at 1 GHz), 1mm in radius, and are placed asymmetrically. Antennas 1 and 2 are quarter-wavelength spaced, and antennas 1 and 3 are sixth-wavelength spaced. Antenna 2 and 3 are placed at a right angle from antenna 1. Note that asymmetry is introduced on purpose to test the generality of our proposed DN. The response of the three-port antenna is shown in Figure \ref{fig:monopole_3ports} (b), where mutual couplings around -10 dB are observed. 

\begin{table}
\renewcommand{\arraystretch}{1.5}
\captionsetup{justification=centering}
\caption{Decoupling Network Parameters for the Three-Port System with Standard Electrical Length} 
\centering 
\begin{tabular}{|c |c | c ||c|c|c|} 
\hline
TL Branch & $Z_0$ ($\Omega$) & $\theta$ (degree) & TL Branch & $Z_0$ ($\Omega$) & $\theta$ (degree)\\
\hline 
$TL_{11}$ & 65.75 & 225 & $TL_{33}$ & 317.21 & 135\\
\hline 
$TL_{12}$ & 3492.69 & 135 & $TL_{34}$ & 73.93 & 135\\
\hline 
$TL_{13}$ & 679.90 & 135 & $TL_{35}$ & 55.25 & 135\\
\hline 
$TL_{14}$ & 141.82 & 225 & $TL_{36}$ & 62.07 & 135\\
\hline 
$TL_{15}$ & 371.42 & 135 & $TL_{44}$ & 131.24 & 135\\
\hline 
$TL_{16}$ & 168.33 & 135 & $TL_{45}$ & 249.31 & 135\\
\hline 
$TL_{22}$ & 31.79 & 225 & $TL_{46}$ & 262.15 & 225\\
\hline 
$TL_{23}$ & 164.19 & 225 & $TL_{55}$ & 233.16 & 225\\
\hline 
$TL_{24}$ & 307.39 & 135 & $TL_{56}$ & 94.87 & 135\\
\hline 
$TL_{25}$ & 56.14 & 225 &$TL_{66}$ & 37.90 & 225\\
\hline 
$TL_{26}$ & 86.59 & 135 &  &  & \\
\hline 
\end{tabular}
\label{tab:standard_3port} 
\end{table}

A DN using the standard electrical length is synthesized at 1GHz based on the method discussed in Section \ref{section:math_TL_DN}. Similar to previous procedures, the calcuation starts with the $S_L$ of the antenna at 1 GHz, and then $S_{DN}$ is assembled based on \eqref{eq:svd_DN}. Then solving \eqref{eq:Y_pi_Y_DN} gives the decoupling networks. The resulted TL-DN parameters are given in Table \ref{tab:standard_3port}. Ports 1, 2, 3 corresponds to the decoupled ports, and ports 4, 5, 6 corresponds to the antenna ports. The decoupled three-port response is shown in Figure \ref{fig:monopole_3ports} (c). Again, all three ports are decoupled at the design frequency 1 GHz ($S_{12},S_{13},S_{23}<-50$dB).

Both examples validate the proposed DN. The only design parameters in the synthesized DN are the characteristic impedance of each branch of the $\pi$-network (with the electrical length being standardized). For branches with extremely large characteristic impedance, it can be left out without impacting the performance. It's worth noting that the generalized $\pi$-Network can be further simplified to reduce implementation complexity \cite{nie2014systematic} if so desired.

\section{Conclusions}
\label{sec:Conclusions}
In this paper, we introduced a TL-based decoupling network that achieves decoupling for arbitrarily coupled loads with an arbitrary number of ports at a given design frequency. A standardized electrical length of $3\lambda/8$ or $5\lambda/8$ is picked to facilitate implementation, leaving the characteristic impedances of the transmission line branches as the only design parameters. Two coupled antenna systems are investigated as examples and perfect decoupling at the design frequencies are achieved, validating the proposed decoupling network. The proposed DN is solely based on transmission lines, which can be precisely engineered for performance consistency and can be easily scaled for antennas operating at different frequencies.

\bibliographystyle{IEEEtran}
\bibliography{DN}

\end{document}